# Instantaneous cell migration velocity may be ill-defined

Gilberto L. Thomas[1*], Ismael Fortuna[1], Gabriel C. Perrone[1], James A. Glazier[2,3], Julio M. Belmonte[2,4], and Rita M.C. de Almeida[1,5*]

[1]*Instituto de Física* and [5]*Instituto Nacional de Ciência e Tecnologia: Sistemas Complexos, Universidade Federal do Rio Grande do Sul, Porto Alegre, RS, Brazil*
[2]*Biocomplexity Institute* and [3]*Department of Intelligent Systems Engineering, Indiana University, Bloomington, Indiana, United States of America*
[4]*Department of Physics, North Carolina State University, North Carolina, United States of America*

*Corresponding authors: glt@if.ufrgs.br, rita@if.ufrgs.br

Cell crawling is critical to biological development, homeostasis and disease. In many cases, cell trajectories are quasi-random-walk. *In vitro* assays on flat surfaces often described such quasi-random-walk cell trajectories as approximations to a solution of a Langevin process. However, experiments show quasi-diffusive behavior at small timescales, indicating that instantaneous velocity and velocity autocorrelations are not well-defined. We propose to characterize mean-squared cell displacement using a modified Fürth equation with three temporal and spatial regimes: short- and long-time/range diffusion and intermediate time/range ballistic motion. This analysis collapses mean-squared displacements of previously published experimental data onto a single-parameter family of curves, allowing direct comparison between movement in different cell types, and between experiments and numerical simulations. Our method also show that robust cell-motility quantification requires an experiment with a maximum interval between images of a few percent of the cell-motion persistence time or less, and a duration of a few orders-of-magnitude longer than the cell-motion persistence time or more.

Cell crawling and migration is ubiquitous in biological processes such as embryonic development [1], wound healing [2], inflammatory response [3] and many pathologies [4]. Cell motion often correlates with cell polarization, with strong positive feedback between cell movement and biochemical and structural polarization within the cell [5,6,7,8,9].

Experiments usually quantify cell motion from time-series images (movies) of the migrating cells. Center of mass trajectories of cells often have large quasi-random-walk components, inspiring characterization of trajectories in terms of the statistics of random walks. However, many experiments either set intervals between images too long or experiment durations too short to allow robust quantification of cell migration statistical properties.

Here, we present a method for robust quantification of quasi-random-walk cell migration. We start with a modified Fürth equation for the mean-squared displacement (**MSD**) that includes three temporal regimes: short-time/range and long-time/range diffusion and mid-time/range quasi-ballistic motion. We show that fitting for the three parameters of the modified Fürth equation and its second derivative allows rescaling of time and length to collapse five disparate sets of experimental MSD plots onto a single-parameter family of curves. The modified Fürth equation determines the maximum time interval between images and minimum experiment duration required for robust characterization of cell movement.

The Fürth-equation analysis defines the natural time and length rescaling to employ to allow proper comparison between experiments. It also reveals a problem with cell velocity measurements. Cells' small-time-scale diffusive behavior means that the ratio of displacement over time interval diverges as the time interval goes to zero; that is, instantaneous velocity is not well defined, but depends on the chosen time interval. We show that this effect is present, though often overlooked, in experiments. The definition of the velocity autocorrelation also appears problematic. However, we show that the autocorrelation of the displacement over a finite time interval is well-defined, provided that we choose the time interval carefully.

*Cell-motion quantification.* The modulus and direction of the velocity suffice to quantify ballistic movement with constant velocity. For normal diffusive motion we can use the MSD of the cell's center of mass, $\langle |\Delta \vec{r}|^2 \rangle$, defined as:

$$\langle |\Delta \vec{r}|^2 \rangle = \langle \frac{1}{T-\Delta t} \int_0^{T-\Delta t} dt (\vec{r}(t+\Delta t) - \vec{r}(t))^2 \rangle, \quad (1)$$

where the time integral extends over the experiment duration $T$, and $\langle \cdot \rangle$ stands for averages over different experiments or different cells of the same type within an experiment. The integral averages over the whole experiment duration, provided $t < T - \Delta t$. For normal diffusion, $\langle |\Delta \vec{r}|^2 \rangle \sim \Delta t$, with the slope of $\langle |\Delta \vec{r}|^2 \rangle$ versus $\Delta t$ defining the *diffusion coefficient D*. However, when cell displacements are neither purely ballistic or diffusive, we cannot quantify their movement by considering $\vec{v}$ or $D$ alone.

A persistent random walk (**PRW**) interpolates between ballistic and diffusive movement and is a common model for the movement of cells in *in vitro* assays in 2 and 3 dimensions. The canonical model for a persistent random walk is a Langevin equation with white noise, also known as an Ornstein-Uhlenbeck process [10]. The resulting MSD equation in 2 dimensions is:

$$\langle |\Delta \vec{r}|^2 \rangle = 4D \left( \Delta t - P(1 - e^{-\Delta t/P}) \right), \quad (2)$$

where the *persistence time*, $P$, sets the timescale of the transition between short time ($\Delta t \ll P$) ballistic motion and long time ($\Delta t \gg P$) diffusive motion with diffusion coefficient $D$. Fits to $D$ and $P$, as proposed by Fürth [11], agree well with experimentally-observed cell trajectories for certain cell types [12]. As $\langle |\Delta \vec{r}|^2 \rangle \sim 4D(\Delta t)^2/P$ for $\Delta t \to 0$, instantaneous velocity is well-defined.

However, cell trajectories frequently deviate from Fürth behavior [13,14,15,16,17,18] with slower movement for small $\Delta t$ and at least two apparent time scales in the velocity auto-correlation functions. The Ornstein-Uhlenbeck process has only one time-scale [13,14]. Dieterich and collaborators observed this deviation and, to fit experimental cell trajectory data at short time intervals, included an *ad-hoc* noise term in the mean-squared displacement solution [14].

Here, we follow a different path. To describe the short-time interval regime, we augment the Fürth equation with third, short-time diffusive regime:

$$\langle |\Delta \vec{r}|^2 \rangle = 4D \left( \frac{\Delta t}{1-S} - P(1 - e^{-\Delta t/P}) \right), \quad (3)$$

where $0 \leq S < 1$ is the fraction of the persistence time $P$ at which the short-time diffusive behavior ends and cell movement becomes ballistic (see Fig. S1 in supplementary materials [19]). We may rescale eq. (3) in terms of natural units [19], *i.e.*, time in terms of the persistence-time-scale $P$ and length in terms of the persistence length-scale $\sqrt{4DP/(1-S)}$, by defining $\tau \equiv t/P$ and $\vec{\rho} \equiv \vec{r}/\sqrt{4DP/(1-S)}$:

$$\langle |\Delta \vec{\rho}|^2 \rangle = \Delta \tau - (1-S)(1 - e^{-\Delta \tau}), \quad (4)$$

leaving a family of curves specified by a **single** dimensionless parameter, $0 \leq S < 1$, that gives the relative duration of the short-time diffusive regime. However, now $|\Delta \vec{\rho}|^2 \sim S \Delta \tau$ for $\Delta \tau \to 0$ and instantaneous velocity is not well-defined.

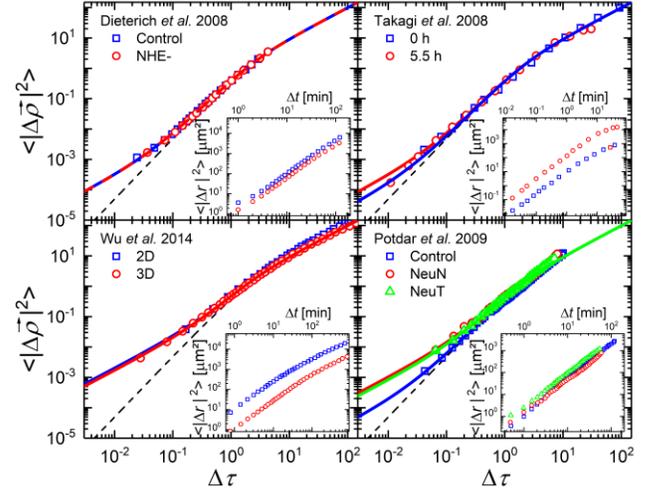

*Figure 1 MSD curves for different cells, treatments, and laboratories, as indicated, fitted using Eq. 4 in rescaled, dedimensionalized units, with the raw measurements shown in the insets. For more details, see Table 1, below.*

*Comparison with experimental data.* To compare the original Fürth equation with our modified form, we fit both models (and their second derivative) to several published experimental trajectories for isolated cells crawling on different substrates [14,15,20,21]. For details see supplementary materials online [19]. Figure 1 shows the fitted curves and Table 1 provides details on the experiments and the fitted values for $D$, $P$, and $S$. For mid- and long-range time intervals, both models fit the experimental data equally well, showing that Eq. (4) suffices to describe the long-time behavior of many types of cell migration. However, the modified Fürth equation fits the data at shorter time intervals (when available), while the original Fürth model does not agree in this regime.

The quality of the fits between experiment and the modified Fürth equation depends on whether the data include information on sufficiently short and long time intervals to capture all three temporal regimes. The smallest timescale in a time-series is the interval between the acquired images and the longest timescale, the duration of the experiment.

The velocity auto-correlation function (**VACF**) provides information on the mechanisms of cell migration:

$$VACF = \langle \frac{1}{T-\Delta t}\int_0^{T-\Delta t} dt\, \vec{v}(t) \cdot \vec{v}(t+\Delta t)\rangle, \quad (6)$$

where $\langle \cdot \rangle$ represents averages over experiments. For a classic persistent random walk, we can obtain the VACF as the second derivative of the MSD curve. When short-time motion is diffusive, the short-time velocity is not well-defined. On the other hand, we can define the average velocity over a finite time interval $\delta$, $\vec{v}_{av}(t,\delta) = \frac{\vec{r}(t+\delta)-\vec{r}(t)}{\delta}$, and then define the average velocity auto-correlation $\psi_\delta(\Delta t)$:

$$\psi_\delta(\Delta t) = \langle \frac{1}{T-\Delta t}\int_0^{T-\Delta t} dt\, \vec{v}_{av}(t,\delta) \cdot \vec{v}_{av}(t+\Delta t,\delta)\rangle. \quad (7)$$

$\psi_\delta$ detects trivial correlations when $\delta < \Delta t$, since the intervals used to calculate $\vec{v}_{av}(t,\delta)$ and $\vec{v}_{av}(t+\Delta t,\delta)$ overlap.

We found the best agreement between experimental MSD and the modified Fürth equation using the data from the experiment by Takagi *et al.* [21] (middle, right panel, Fig. 1). In this experiment, the time interval between images is 1 s and the experiment duration is 2400 s. This range of time scales fully sampled all three temporal regimes. The minimum value of $S$ is on the order of 0.01. Based on these observations, an informative experimental time series requires a time-interval between images $\leq 0.01P$ and a total observation time $\geq 50P$, to explore all three regimes and allow accurate quantification of the modified Fürth equation parameters $D$, $P$ and $S$.

The origin of the short-time diffusive behavior requires discussion. It could be a biologically-significant result of the sub-cellular dynamics of actin and/or it could be an artifact of segmentation noise.

Experiments usually determine a cell's center-of-mass position from an image of the cell projected onto the substrate. Hence, the position of the perimeter of the cell image has an uncertainty of at least $\pm 0.5$ pixel length at each location. The contribution of these uncertainties to the error in the center-of-mass position (average of all pixel positions) is:

$$\varepsilon \sim \left(16\pi \times N \times \left(\frac{R}{\ell}\right)^3\right)^{-1/2}, \quad (5)$$

where $\ell$ is the size of an image pixel in $\mu$m, $R$ is the typical size of the cell nucleus in $\mu$m, and $N$ is the number of displacement measurements used to calculate the MSD. For the first point of the MSD curve $N = (T - \Delta t)/\Delta t$, where $T$ is the duration of the experiment and $\Delta t$ is the interval between measurements. For the experimental data of Takagi *et al.*, we estimate the error of the center-of-mass position for their smallest time interval ($T = 40$ min, $\Delta t = 1$ s) to be $\varepsilon \sim 1.15 \times 10^{-3}$ $\mu$m, where we assume $R/\ell \sim 5$ pixels and a pixel edge length of 1 $\mu$m. This error is much smaller than the square root of the MSD in their two data sets for $\Delta t = 1$ s, *i.e.* 0.35 $\mu$m and 0.128 $\mu$m (see upper right panel, Fig.1). Dieterich *et al.* [14] have shown in more detail that errors in single measurements of cells' positions cannot generally explain the observed diffusive behavior for short time intervals in MSD curves. Together these results indicate that the short-time diffusive behavior is biologically significant, not an experimental artifact.

| Reference | Description | Cell radius | Experiment | D ($\mu$m²/min) | P (min) | S |
|---|---|---|---|---|---|---|
| Potdar et al., 2009 [15] | MCF-10A cell line expressing different versions of promigratory tyrosine receptor Her2/neu; 2D culture on plastic substrate. Time interval between photos: 0.5 min | 5 - 10 $\mu$m | Control | 4.63 | 11.92 | 0.015 |
| | | | neu-N (pre-malignant) | 2.00 | 7.50 | 0.100 |
| | | | neu-T (invasive) | 4.37 | 7.69 | 0.080 |
| Metzner et al., 2015 [20] | MDA-MB-231 cell line; 2D cultures on plastic, fibronectin, and collagen coated substrates. Time interval between photos: 1 min | 5 - 10 $\mu$m | Plastic | 2.15 | 13.41 | 0.206 |
| | | | Fibronectin | 6.69 | 19.41 | 0.020 |
| | | | Collagen | 1.70 | 79.11 | 0.298 |
| Dieterich et al., 2008 [14] | MDCK-F cell line; wild type and NHE-deficient (relevant for cell migration); 2D cultures on plastic substrate. Time interval between photos: 1 min | 5 - 10 $\mu$m | Control | 8.43 | 27.60 | 0.028 |
| | | | NHE-deficient | 18.27 | 41.72 | 0.028 |
| Takagi et al., 2008 [21] | *Dictyostelium discoideum*, prepared in the vegetative state. Data acquired 0 h and 5.5 h after preparation, in the development phase. Time interval between photos: 1 min | 3 - 5 $\mu$m | Vegetative 0 h after preparation | 4.78 | 0.41 | 0.028 |
| | | | 5.5 h after preparation | 123.42 | 1.51 | 0.013 |
| Wu et al., 2014 [17] | WT fibrosarcoma HT1080 cells, over a flat collagen coated dish (2D) and cells embedded in a collagen matrix (3D). Time interval between photos: 2 min | 5 - 10 $\mu$m | 2D | 4.95 | 1.62 | 0.200 |
| | | | 3D | 0.80 | 9.81 | 0.169 |

*Table 1 Details and references for the analyzed experimental data.*

Figure 2 analyzes three sets of experiments by Metzner [20]. The upper left panel shows the average speed $\langle |\vec{v}_{av}(\tau,\delta)|\rangle$, as a function of $\delta$ (in natural units): this quantity does not converge to a finite value as $\delta \to 0$, due to the diffusive behavior of the cells over short time intervals. Fig. 2 also computes $\psi_\delta(\Delta \tau)$ directly from the trajectories for different values of $\delta$, and compares the second derivatives of Eq. (4) and the MSD experimental data. All derived quantities agree for the three experimental sets for $\Delta \tau > S$, indicating a stationary process. For $S > \Delta \tau > \delta$, $\psi_\delta(\Delta \tau)$ and the second derivative of the experimental MSD decrease, indicating a loss of memory by the velocity, typical of diffusive behavior. The analytical second derivative of Eq. (4) does not fit the experimental data for small time intervals. Together, the behavior of $\langle |\vec{v}_{av}(\tau,\delta)|\rangle$ and $\psi_\delta(\Delta \tau)$

provide strong evidence that cell migration at small time scales is diffusive.

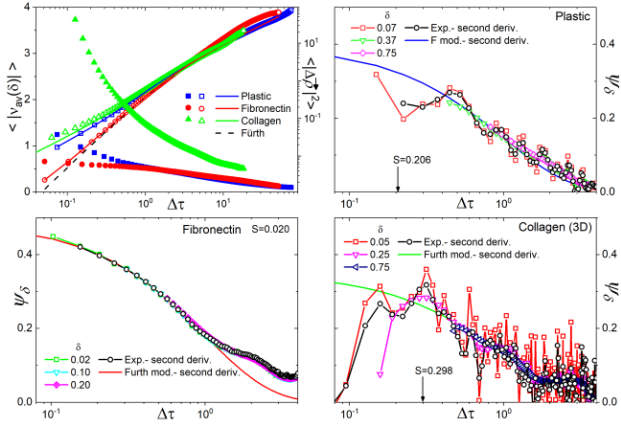

Figure 2. Analysis of Metzner et al. experiments. Top left panel: average speed as a function of δ, together with MSD curves for three different experiments. Other panels: Average $\psi_\delta(\Delta\tau)$, as a function of $\Delta\tau$ for different values of δ for three sets of experiments.

Figure 3 present some of the experimental trajectories for cells crawling on plastic by Metzner *et al.* (Figs. S2 and S3 in the supplementary materials analyze cell trajectories on collagen and fibronectin [19]). The trajectories in Fig. 3 clearly show small-length-scale behavior compatible with quasi-diffusive motion at small time scales.

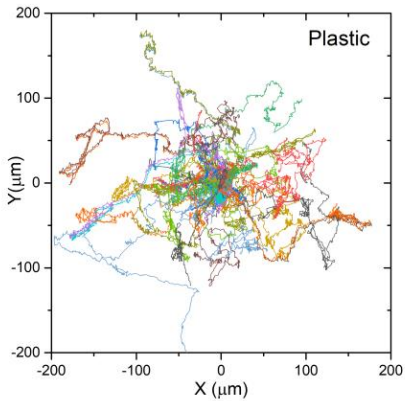

Figure 3. Trajectories from Metzner experiment for cells crawling on plastic. See Table 1 for details.

*Discussion and Conclusion.* Cell-migration experiments typically track center-of-mass positions of individual cells to obtain MSD *vs*. time interval. Comparison between experiments, however, requires rescaling of time and length between experiments. The modified Furth equation provides the values for this rescaling. Unlike the classic Fürth equation, it also describes the short-time/range diffusive motion usually observed in experiments. Many authors either attribute this short-time/range diffusive motion to measurement error, or simply disregard it. This oversight is unfortunate, because the short-time/range diffusive regime provides quantitative information on the sub-cellular mechanisms that generate cell motility and determine the time a leading edge takes to form, dissipate or reorganize to destabilize the cell polarization responsible for the ballistic regime, and hence may be critical to explain the mechanism of cell migration for a specific cell type. Furthermore, when the short time motion is diffusive, instantaneous velocity is not well-defined and measurements of both velocity and velocity autocorrelation require the definition of an average velocity and associated autocorrelation function. In this case, results may depend on the time interval δ used to calculate average velocity.

The results presented here also provides a possible answer to a long-standing paradox in cell migration: while cell polarization appears to determine cell migration direction, demonstrating the correlation between cell velocity and cell polarization has been difficult: for this analysis to work, the direction of cell movement and average velocity must be measured over time intervals $S < \Delta\tau < 1$ in the intermediate, quasi-ballistic regime.

Finally, we point that trajectories obeying the modified Furth equation emerge from dynamic equations, analogous to the Langevin problem. We will address this point in future work.

RMCdeA and GLT acknowledge the hospitality of the Biocomplexity Institute, Indiana University, Bloomington and financial support from Brazilian agencies CNPq, CAPES (code 001), FAPERGS, and from FAPERGS-PRONEX grant 10/0008-0. RMCdeA and JAG acknowledge support from the Falk Medical Research Trust Catalyst Program and the US National Institutes of Health, grants U01 GM111243, R01 GM076692 and R01 GM122424 and National Science Foundation grant 1720625.

# Supplementary material for

# Instantaneous cell migration velocity may be ill-defined

Gilberto L. Thomas[1*], Ismael Fortuna[1], Gabriel C. Perrone[1], James A. Glazier[2,3], Julio M. Belmonte[2,4], and Rita M.C. de Almeida[1,5*]

[1]Instituto de Física and [5]Instituto Nacional de Ciência e Tecnologia: Sistemas Complexos,
Universidade Federal do Rio Grande do Sul, Porto Alegre, RS, Brazil
[2]Biocomplexity Institute and [3]Department of Intelligent Systems Engineering, Indiana University, Bloomington, Indiana, United States of America
[4]Department of Physics, North Carolina State University, North Carolina, United States of America

**Modified Fürth equation and natural units**.

The modified Fürth equation is:

$$\langle |\Delta \vec{r}|^2 \rangle = 4D \left( \frac{\Delta t}{1-S} - P(1 - e^{-\Delta t/P}) \right). \quad (S1)$$

Eq. S1 reduces to classical Fürth equation when $S = 0$. To show the scaling more naturally, we can rewrite eq. S1:

$$\frac{\langle |\Delta \vec{r}|^2 \rangle}{\left( \frac{4DP}{1-S} \right)} = \frac{\Delta t}{P} - (1-S)(1 - e^{-\Delta t/P}), \quad (S2)$$

Suggesting that $\sqrt{\frac{4DP}{1-S}}$ is a natural length-scale to quantify movement and $P$ a natural time scale. If we definte the dimensionless length and time as $|\Delta \vec{\rho}| \equiv \frac{|\Delta \vec{r}|}{\sqrt{\frac{4DP}{1-S}}}$ and $\Delta \tau \equiv \frac{\Delta t}{P}$, the modified Fürth equation simplifies to:

$$\langle |\Delta \vec{\rho}|^2 \rangle = \Delta \tau - (1-S)(1 - e^{-\Delta \tau}). \quad (S3)$$

Equation S3 thus defines a single-parameter family of curves.

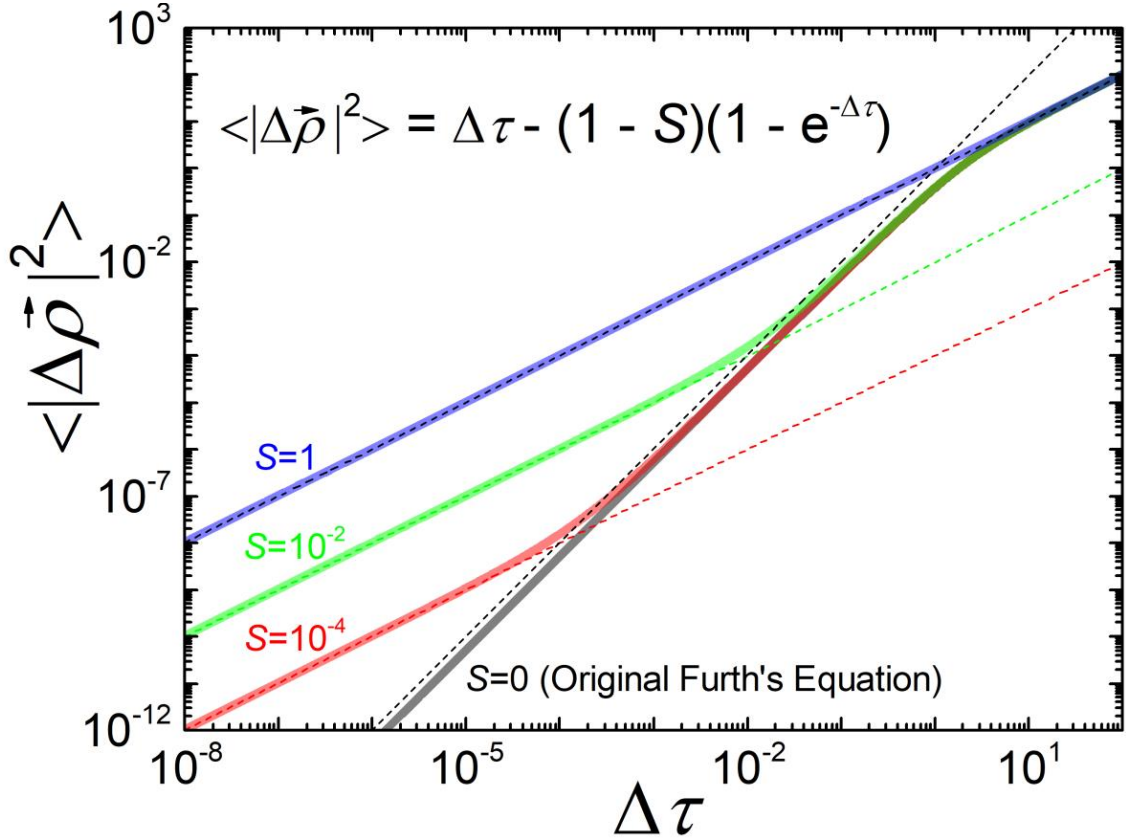

Figure S1. Plot of the modified Fürth equation, showing diffusive behavior for small time intervals, and ballistic and diffusive behaviors for longer time intervals. $\Delta\vec{\rho}$ and $\Delta\tau$ are, respectively, displacement and time, both rescaled by the persistence time-scale $P$ and persistence length-scale $\sqrt{4DP/(1-S)}$. $S$, the only free dimensionless parameter of the model, defines the time-scale of the transition from short-time diffusive movement to ballistic movement. The rescaled $\Delta\tau = 1$ corresponds to the unscaled $\Delta t = P$.

**Fitting procedure to determine $P$, $D$, and $S$**

Data describing a cell's trajectory consists of successive values for the coordinates $x$, $y$ (and $z$ for 3D trajectories), describing the position of the cell measured at fixed time intervals. To determine the values of $P$, $D$ and $S$ from such a time-series we follow these steps:

1- Caculate the mean square displacement (MSD) as a function of time interval from the experimental time series.

2- Calculate the numerical second derivative of this curve (which is just two times the value of the velocity autocorrelation function).

3- Calculate the second derivative of the modified Fürth equation (eq.(S1)):

$$\frac{d^2}{d(\Delta t)^2}\langle|\Delta\vec{r}|^2\rangle = \frac{4D}{P}e^{-\Delta t/P}. \qquad (S4)$$

Observe that this second derivative (eq. S4) does not depend on $S$.

4-  Use eq. (S4) to fit the experimental second derivative of the MSD curve and obtain the values of $D$ and $P$.

5-  With these values, fit the experimental MSD curve to eq. (S1) to obtain $S$.

6-  With $D$, $P$, and $S$, we can then calculate the natural length and time scales $\sqrt{\frac{4DP}{1-S}}$ and $P$ to dedimensionalize the experimental length and time measurements.

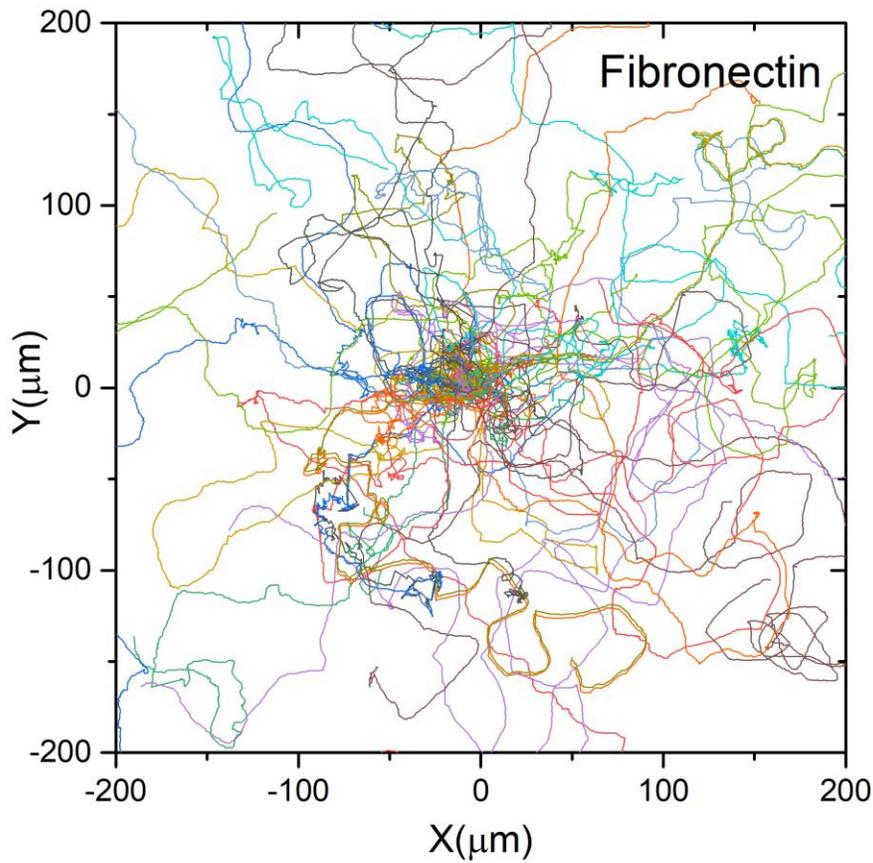

Figure S2. Experimental trajectories for cells crawling on fibronectin by Metzner [1]. Metzner measured the trajectories using time intervals between snapshots on the order of the time-scale $S$ found for these cells. Consequently, most of the trajectories do not show the expected short-time diffusive motion of the cells.

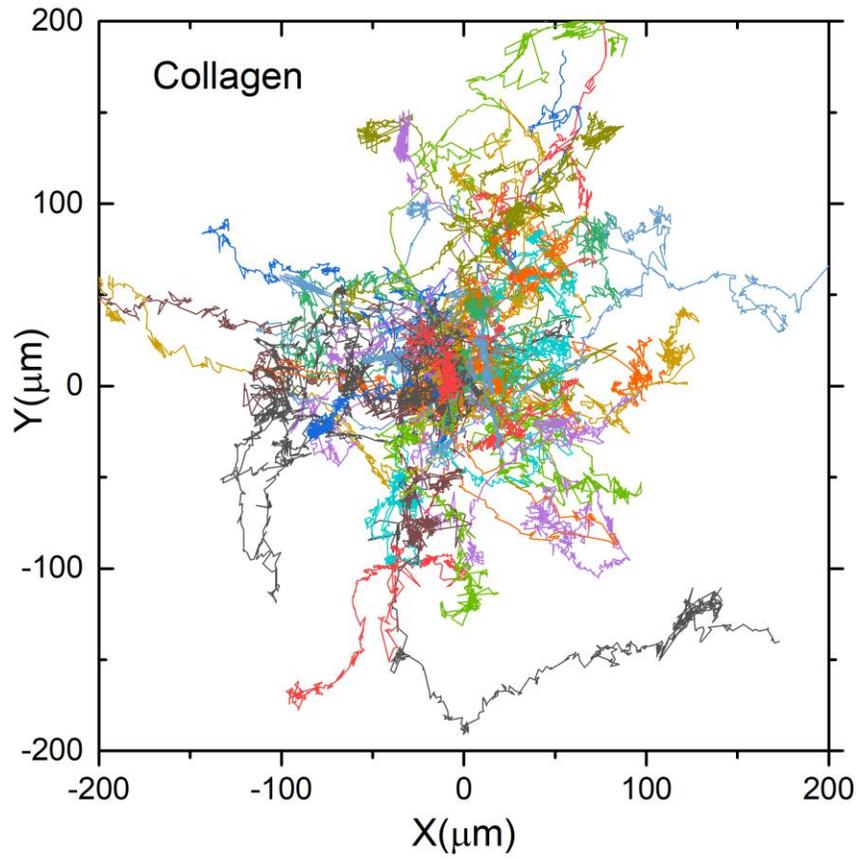

Figure S3. 2D projection of experimental 3D trajectories for cells crawling in collagen by Metzner [1]. The trajectories clearly show small-length scale behavior compatible with quasi-diffusive motion over short time scales.

[1] C. Metzner, C. Mark, J. Steinwachs, L. Lautscham, F. Stadler, and B. Fabry, Nat Commun **6**, 7516 (2015).